%% file: main.tex
\documentclass[sigconf]{acmart}

\usepackage{booktabs} % For formal tables
\usepackage{multirow}
\usepackage{subfigure}
\usepackage{verbatim}
\usepackage{CJKutf8}
\usepackage{flushend}

\usepackage{makecell}
\usepackage[utf8]{inputenc}
\copyrightyear{2022}
\acmYear{2022}
\setcopyright{acmcopyright}
\acmConference[KDD '22]{Proceedings of the 28th SIGKDD Conference on Knowledge Discovery and Data Mining}{August 14--18, 2022}{Washington D.C.}
\acmBooktitle{Proceedings of the 28th SIGKDD Conference on Knowledge Discovery and Data Mining (KDD '22), August 14--18, 2022, Washington D.C.}
\acmPrice{15.00}
\begin{document}

\title{FedAttack: Effective and Covert Poisoning Attack on Federated Recommendation via Hard Sampling}

\fancyhead{}

%\subtitle{FedCTR}
%\subtitlenote{The full version of the author's guide is available as
%  \texttt{acmart.pdf} document}

\author{Chuhan Wu$^1$, Fangzhao Wu$^2$, Tao Qi$^1$, Yongfeng Huang$^1$, Xing Xie$^2$}

\affiliation{%
  \institution{$^1$Department of Electronic Engineering, Tsinghua University, Beijing 100084 \\ $^2$Microsoft Research Asia, Beijing 100080, China}
} 
\email{{wuchuhan15,wufangzhao,taoqi.qt}@gmail.com,yfhuang@tsinghua.edu.cn,xingx@microsoft.com}

% \author{Chuhan Wu}
% \affiliation{%
%     \city{Beijing}
%     \postcode{100084}
%   \state{China}
% }
% \email{wuchuhan15@gmail.com}

% \author{Fangzhao Wu}
% \affiliation{%
%   \institution{Microsoft Research Asia}
%   \city{Beijing}
%   \state{China}
%   \postcode{100080}
% }
% \email{wufangzhao@gmail.com}

% \author{Tao Qi}
% \affiliation{%
%     \city{Beijing}
%     \postcode{100084}
%   \state{China}
% }
% \email{qit16@mails.tsinghua.edu.cn}
% \author{Heyuan Wang}
% \affiliation{%
%   \institution{Microsoft Research Asia}
%   \city{Beijing}
%   \state{China}
%   \postcode{100080}
% }
% \email{heyuanww@163.com}

% \author{Yongfeng Huang}
% \affiliation{%
%     \city{Beijing}
%     \postcode{100084}
%   \state{China}
% }
% \email{yfhuang@tsinghua.edu.cn}

% \author{Xing Xie}
% \affiliation{%
%   \institution{Microsoft Research Asia}
%   \city{Beijing}
%   \state{China}
%   \postcode{100080}
% }
% \email{xing.xie@microsoft.com}

\begin{abstract}
Federated learning (FL) is a feasible technique to learn personalized recommendation models from decentralized user data.
Unfortunately, federated recommender systems are vulnerable to poisoning attacks by malicious clients.
Existing recommender system poisoning methods mainly focus on promoting the recommendation chances of target items due to financial incentives.
In fact, in real-world scenarios, the attacker may also attempt to degrade the overall performance of recommender systems.
However, existing general FL poisoning methods for degrading model performance are either ineffective or not concealed in poisoning federated recommender systems.
In this paper, we propose a simple yet effective and covert poisoning attack method on federated recommendation, named \textit{FedAttack}.
Its core idea is using globally hardest samples to subvert model training.
More specifically, the malicious clients first infer user embeddings based on local user profiles.
Next, they choose the candidate items that are most relevant to the user embeddings as hardest negative samples, and find the candidates farthest from the user embeddings as hardest positive samples.
The model gradients inferred from these poisoned samples are then uploaded to the server for aggregation and model update.
Since the behaviors of malicious clients are somewhat similar to users with diverse interests, they cannot be effectively distinguished from normal clients by the server.
Extensive experiments on two benchmark datasets show that \textit{FedAttack} can effectively degrade the performance of various federated recommender systems, meanwhile cannot be effectively detected nor defended by many existing methods.

\end{abstract}



\maketitle

\input{data/introduction.tex}

\input{data/relatedwork.tex}

\input{data/method.tex}
\input{data/experiment.tex}

\input{data/conclusion.tex}

\bibliographystyle{ACM-Reference-Format}
\bibliography{main}

\end{document}

%% file: data/introduction.tex
\section{Introduction}

Personalized recommendation techniques are widely used in many online scenarios such as e-commerce~\cite{kang2018self} and news feed~\cite{wu2019npa} to alleviate users' information overload.
Conventional recommender systems are learned on centralized user data, which may arouse considerable privacy concerns and risks~\cite{yang2020federated}.
Federated learning (FL)~\cite{mcmahan2017communication} is a privacy-preserving paradigm that enables learning models on decentralized data under privacy protection~\cite{yang2019federated}.
It has been successfully applied to the recommendation field to train privacy-aware recommender systems without collecting and exchanging raw user data~\cite{lin2020fedrec,wu2021fedgnn,liang2021fedrec}, where user privacy can be protected to a certain extent.

Unfortunately, due to the decentralization of model training, not all clients in federated learning are trusted and a fraction of them may be controlled by malicious parties~\cite{tolpegin2020data}.
Thus, federated learning-based systems are more vulnerable to poisoning attack than centralized learning~\cite{bagdasaryan2020backdoor}.
In federated recommendation, most prior poisoning works focus on promoting target items by increasing their exposure chances~\cite{zhang2021pipattack}, which is usually motivated by financial incentives~\cite{chen2020robust}.
Besides item promotion, the attacker may aim to degrade the overall performance of federated recommender systems~\cite{lyu2020privacy}, which may be due to the cut-throat competitions among companies~\cite{mahajan2015malicious}.
This attack form is known as untargeted attack~\cite{lyu2020threats}.
In this paper, we study untargeted poisoning attack on federated recommendation, which is a less-explored  problem.

\begin{figure}[!t]
  \centering
    \includegraphics[width=1.0\linewidth]{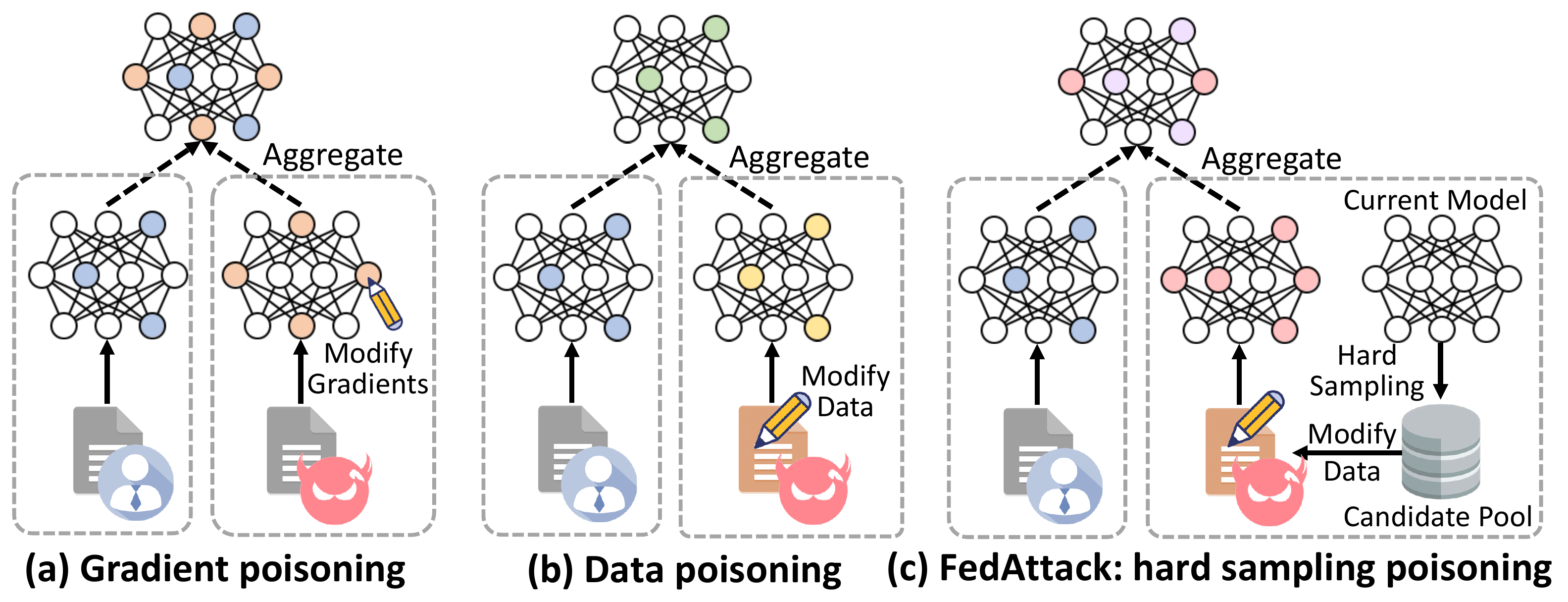}
  \caption{Existing popular poisoning paradigms on FL and our \textit{FedAttack} for poisoning federated recommendation.}
  \label{fig.exp}
\end{figure}

There are several methods for untargeted poisoning attack on federated learning in general domains~\cite{baruch2019little,fang2020local}.
One way is poisoning model gradients, as shown in Fig.~\ref{fig.exp}(a).
A naive method is adding strong noise to the local model gradients uploaded to the server for aggregation.
However, the outlier gradients can be easily recognized by  anomaly detection methods~\cite{nguyen2019diot} or defended by robust gradient aggregation methods~\cite{blanchard2017machine}.
Thus, a few studies explore more covert ways to poison the model gradients. 
For example, \citet{baruch2019little} proposed to add a small amount of noise to each dimension of the average model updates from benign clients, where the noise intensity is estimated by the number of benign and Byzantine clients.
\citet{fang2020local} proposed to perturb the gradients by adding noise with opposite parameter update directions inferred from benign updates.
However, these methods usually require a large fraction of Byzantine workers (e.g., 20\%) to achieve a notable performance decrease, which is quite expensive when there are a large number of clients in FL~\cite{shejwalkar2022back}.
In addition, they need strong assumptions on the attackers' capability such as background knowledge of benign clients' updates and the server's aggregation rule, which may be somewhat difficult to obtain in realistic attacking scenarios.
Another possible way for untargeted attack is data poisoning, as shown in Fig.~\ref{fig.exp}(b).
Most prior studies on data poisoning are conducted in centralized settings~\cite{schwarzschild2021just}.
Some methods aim to generate adversarial samples~\cite{yang2017generative,munoz2019poisoning,chen2019data}, which usually require strong background knowledge of training data.
However, they are not applicable in federated learning because the attackers cannot access other clients' data.
Label flipping is a simple data poisoning method without prior knowledge of training data~\cite{xiao2012adversarial}.
For example, \citet{fang2020local} showed the possibility of poisoning federated learning models by changing true labels to false labels on each Byzantine client.
However, label flipping is also ineffective when the ratio of controlled clients is insufficient~\cite{shejwalkar2022back}.

Different from poisoning general FL systems, untargeted poisoning attack on federated recommendation faces several unique challenges.
First, the number of clients participating in federated recommendation can be rather large, and it can be extremely expensive to control a significant portion of clients~\cite{qi2020privacy}.
Thus, the attacking algorithm needs to be effective under limited Byzantine workers.
Second, many recommendation models are learned on implicit user feedback that contains heavy noise~\cite{wang2021denoising}, and thereby the models are robust to perturbation to some extent~\cite{yu2020sampler}.
Thus, noise injecting and label flipping operations may have very limited impacts on final recommendation performance.
Third, since local user data usually only involves a small subset of items, the updates of local parameters, especially item embeddings, can be very sparse~\cite{bhargava2015and}.
Therefore, aggressive modification on local model updates may frequently trigger anomaly detection and defense mechanisms on the aggregation server.

To address the above challenges, in this paper, we propose a both effective and covert poisoning attack method named \textit{FedAttack}\footnote{Source code is available at https://github.com/wuch15/FedAttack}, which aims to degrade the performance of federated recommender systems with a small fraction of malicious clients.
The core idea of our approach is inspired by hard negative sampling~\cite{kalantidis2020hard}, which is a widely used technique in information retrieval to find informative negative samples.
Although hard samples are beneficial for learning more discriminative models~\cite{robinson2020contrastive}, globally hardest negative examples are usually harmful to model performance because many of them are false negatives and they can lead to bad local minima early in training~\cite{xuan2020hard}.
Thus, in our work we propose to use globally hardest sampling as a poisoning technique, as shown in Fig.~\ref{fig.exp}(c).
More specifically, on each Byzantine client we first use the current model to infer the user embedding from local user profiles.
Next, each Byzantine client uses its local user embedding to retrieve globally hardest negative samples with the closest distance to the user embedding from the candidate pool.
To further increase the difficulty of model training, we also retrieve pseudo ``hardest positive samples'' that are farthest from user embeddings to replace the original positive samples.
By training recommendation models on these hard samples, the Byzantine clients can finally generate poisoned gradients to upload.
These gradients have significant impacts on model convergence, while hard for the server to perceive because they are similar to the gradients generated by benign users with strong preferences on the diversity of recommendation results~\cite{wu2021news}.\footnote{These users may interact with items that are diverse from previously interacted ones, which have similar effects on model learning with hard sampling. }
Extensive experiments on two benchmark recommendation datasets show that \textit{FedAttack} can effectively degrade the performance of various federated recommendation models with a small fraction of Byzantine clients and can circumvent many defense mechanisms.

The contributions of this paper are listed as follows:
\begin{itemize}
    \item To our best knowledge, our approach is the first untargeted poisoning attack method on federated recommendation. It reveals the security risk of recommenders even under decentralized settings with a small fraction of malicious clients, and can raise researchers' attention on enhancing the security of federated recommender systems.
    \item We are the first to employ hard sampling mechanism as an attack technique for poisoning federated recommendation.
    We demonstrate that globally hardest samples can subvert the model training in an effective and covert way.
    \item We conduct extensive experiments on two benchmark  datasets, validating that \textit{FedAttack} is more effective than many baseline methods in attacking and defense circumventing.
\end{itemize}

%% file: data/relatedwork.tex
\section{Related Work}\label{sec:RelatedWork}

The security issues in machine learning systems have been extensively studied~\cite{chio2018machine}, and poisoning attack is an emerging threat in recent years~\cite{shejwalkar2022back}.
Data poisoning is one of the most common forms of poisoning attack~\cite{steinhardt2017certified}.
With a fraction of designed points injected into the training data, the malicious attacker can degrade the model performance in an indiscriminate or targeted way~\cite{munoz2019poisoning}.
Label flipping is a simple and widely used method without  the need of background knowledge of training data~\cite{xiao2015support}.
For example, \citet{xiao2012adversarial} proposed a label flipping method that aims to change true sample labels into false ones which can maximize the classification error.
However, these carefully designed samples can be  identified as outliers and filtered~\cite{taheri2020defending}.
Another popular data poisoning way is generating adversarial  samples to subvert the training process~\cite{wallace2021concealed}.
For example, \citet{yang2017generative} proposed to use autoencoder as the generator to create poisoned data, which  is further updated based on a reward function of loss.
The target model serves as the discriminator that calculates the loss of discriminating between poisoned data and normal data.
\citet{munoz2019poisoning} proposed a similar generative adversarial network based data poisoning method that uses a generator to generate poisoning samples which can maximize classification errors in the target task and meanwhile minimize the discriminator’s ability in distinguishing poisoned data from normal data.
These data poisoning methods are conducted on centralized datasets, where the attacker has strong prior knowledge of the training dataset~\cite{chen2021pois}.
Thus, they are not applicable when training data is decentralized and cannot be exchanged. 

In recent years, several works study the poisoning attack problem in general federated learning~\cite{cao2019understanding}.
According to the attack goals, they can be classified into targeted attack (misclassify specific subsets of inputs),  backdoor attack (misclassify inputs with certain patterns), and untargeted attack (degrading overall performance)~\cite{shejwalkar2022back}.
Here we focus on untargeted attack that is most relevant to our work.
Some centralized data poisoning methods can be directly applied to FL poisoning.
For instance, \citet{fang2020local} showed the potential of label flipping in poisoning attack on FL by simply replacing true labels with other false ones.
Different from data poisoning in centralized learning, in federated learning the attacker can directly manipulate the model updates to achieve stronger attack performance.
For example, \citet{baruch2019little} proposed to first estimate the standard deviation of benign model gradients, then compute an intensity coefficient based on the number of benign and Byzantine clients, and finally modify the model gradients by adding noise according to the standard deviation and intensity coefficient.
\citet{fang2020local} proposed to perturb the gradients by adding noise with constant values but opposite signs with the  benign model updates.
The attacker needs to empirically select the noise constant  that can effectively circumvent the target defense method.
\citet{shejwalkar2021manipulating} further proposed an improved attacking method by using dynamic and data-dependent malicious direction to perturb the model updates. 
However, these methods usually require a large fraction of Byzantine clients for effective attack, which may lead to heavy costs when there are a large number clients in FL.

Different from poisoning attack on general classification models, recommender system poisoning methods may aim to promote a set of target items due to financial incentives~\cite{chen2019data,song2020poisonrec,zhang2020practical,wu2021triple,zhang2021data}.
For example, \citet{li2016data} proposed to generate malicious samples that optimize the tradeoff between maximizing overall benign recommendation errors and promoting ratings of target items.
\citet{fang2018poisoning} proposed a graph-based recommender system poisoning method that finds optimal edge weights between target items and users as well as assigns maximum ratings to them.
They further added filler items with the largest edge weights and generated random ratings for them to mimic normal users.
These methods are aimed at centralized recommendation model learning, where attackers usually have full knowledge of the training data~\cite{tang2020revisiting}.
In federated recommendation scenarios, the attacker cannot access the data on normal clients, and there are very few methods for federated recommendation poisoning.
PipAttack~\cite{zhang2021pipattack} is a recent item promotion method that promotes the ranking scores and estimated popularity of target items, meanwhile uses a distance constraint that regularizes the distance between poisoned and original gradients.
However, there still lacks research of untargeted poisoning attack on federated recommendation, which has many distinct challenges from general FL poisoning.
Our approach is the first one for untargeted federated recommendation poisoning, which can achieve more effective and covert attacks than many general FL poisoning methods.
It reveals a new form of security risk in federated recommendation that needs to be considered in future recommender system design.

%% file: data/method.tex
\section{Methodology}\label{sec:Model}

We introduce the details of our \textit{FedAttack} approach in this section. 
We first present a definition of the problem studied in this paper, next discuss the prior knowledge and capability of the attacker, then describe the basic recommendation model learning framework on benign clients, and finally introduce how to conduct hard sampling-based poisoning attack on Byzantine clients.

\subsection{Problem Definition}

Assume there are $N$ clients in total for federated recommendation model learning, where $M\%$ of them are Byzantine clients controlled by a malicious attacker.
Each client keeps a local copy of the recommendation model, which is collaboratively learned and updated (note that there may be some personalized parameters on each client like user embeddings).
These clients are all coordinated by a server that receives local model updates from clients and aggregates them into a global update to deliver.
Each client locally maintains a user profile (user ID or historical interacted items) that is used to infer user interest.
On each client, the recommendation model aims to predict a ranking score for each candidate item that indicates the probability of this user interacting with this item.
The items in the candidate set are ranked based on their ranking scores.
The Byzantine clients controlled by the attacker aim to generate poisoned gradients that can effectively degrade the recommendation performance on benign clients, meanwhile cannot be easily detected nor defended by the aggregation server.

\begin{figure*}[!t]
  \centering
    \includegraphics[width=0.9\linewidth]{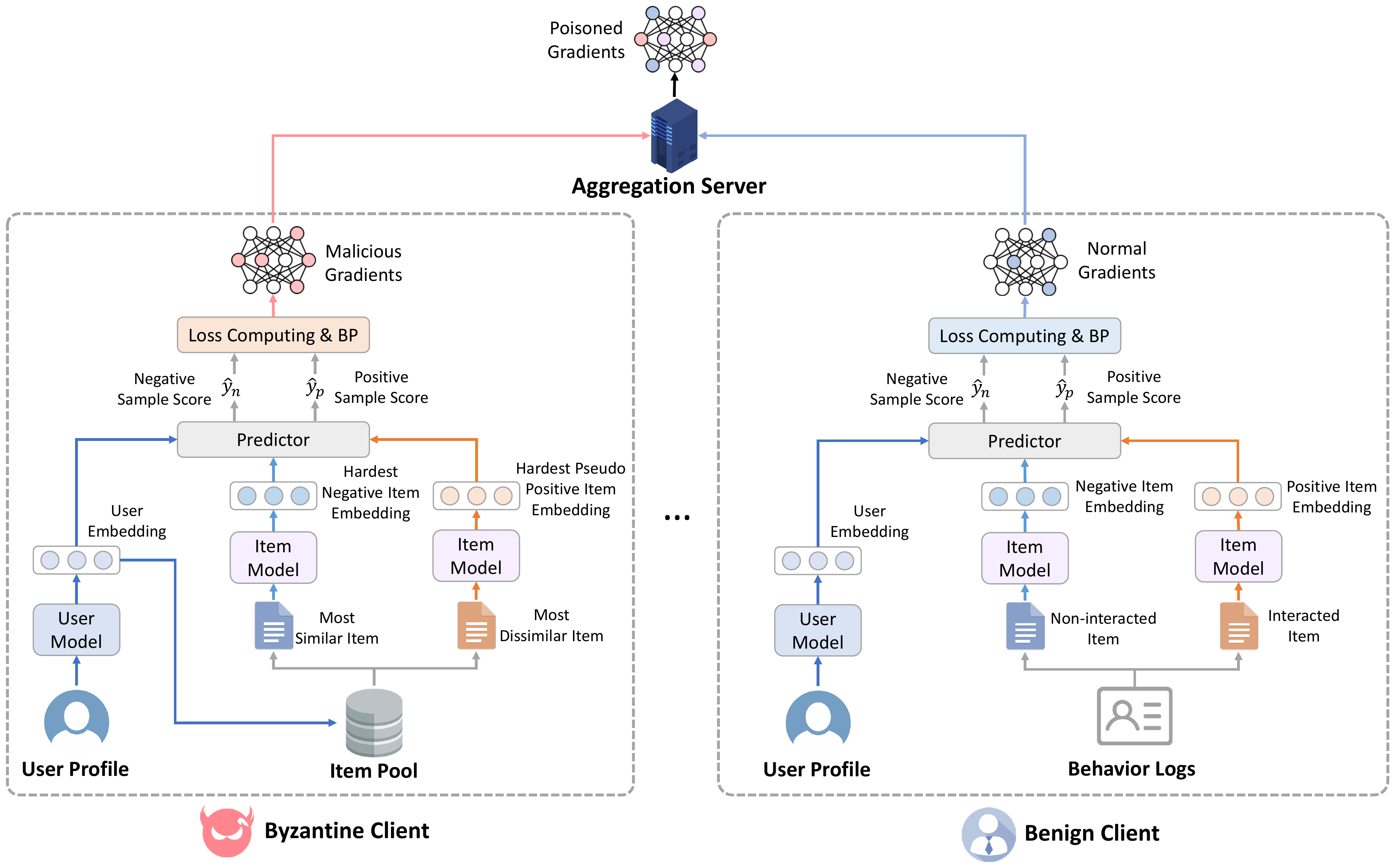}
  \caption{The framework of our \textit{FedAttack} approach.}
  \label{fig.model}
\end{figure*}

\subsection{Assumption and Prior Knowledge}

In \textit{FedAttack}, we assume that the parameter set of the recommendation model includes a user model for inferring user interest embedding from user profiles, an item model that converts item IDs into item embeddings, and a predictor model that computes personalized ranking scores based on the relevance between user interest embeddings and item embeddings.
During the model training, all benign clients only upload their model updates to the server, and Byzantine clients cannot access their raw data or intermediate model results due to privacy reasons.
In addition, the attacker does not know the aggregation rules on the server.
To conduct the attack, we assume that the attacker has the following capability and prior knowledge:
(1) the Byzantine clients have normal user profiles, which can be achieved by hacking normal clients;
(2) the attacker knows a large candidate item pool (would be better if it is the whole item set as~\cite{zhang2021pipattack}), which is achievable because the attacker can crawl the publicly available items on e-commerce platforms;
(3) the Byzantine clients can modify their local model inputs and labels, and have access to their local user and item embeddings (if not satisfied, the attacker can learn a surrogate model to mimic the original model~\cite{zhang2021reverse}).
In our assumptions, the attacker has much weaker capability and  background knowledge of the training data than most existing works on centralized or FL poisoning~\cite{shejwalkar2022back}.
Thus, our attack method is easier to execute in real-world scenarios.

\subsection{FedAttack Framework}

\begin{figure}[!t]
  \centering
    \includegraphics[width=0.92\linewidth]{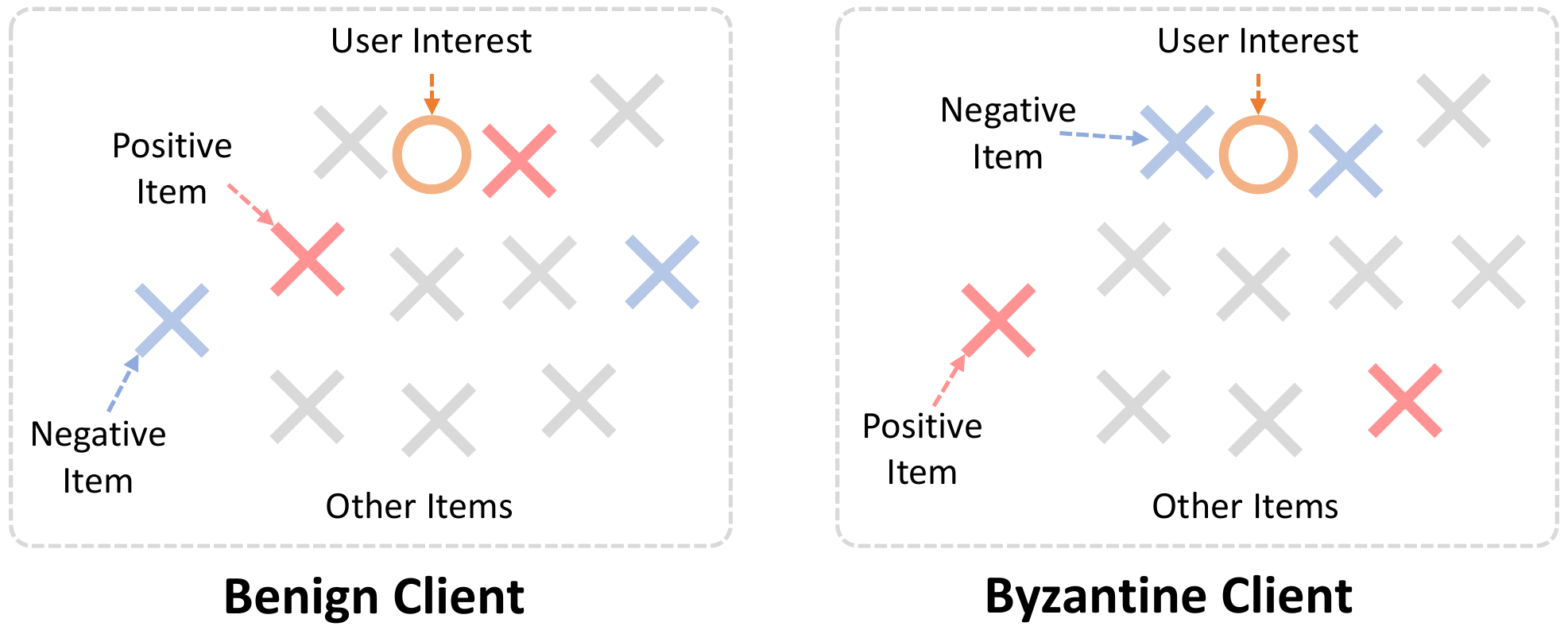}
  \caption{Schematic comparison between negative and positive samples on benign and Byzantine clients.}
  \label{fig.sam}
\end{figure}

The framework of \textit{FedAttack} is shown in Fig.~\ref{fig.model}.
The Byzantine clients only modify the input samples and their labels, while the model gradients are directly computed on these samples without further manipulation.
In each round of model update, the Byzantine clients first use the current user model to infer their user embeddings (denoted as $\mathbf{u}$) from their local user profiles.
The user model can be implemented by various architectures, such as the simple user ID embedding in collaborative filter~\cite{he2017neural}, or sequential models like GRU~\cite{hidasi2015session} and Transformer~\cite{sun2019bert} that learn user embeddings from historical interacted items.
The user embedding is then used to retrieve hard samples on each Byzantine client.
Assume there are $K$ original positive samples on a Byzantine client (can be constructed from the interaction history).
We take BPR~\cite{rendle2009bpr} as an example training scheme to explain the poisoning method as follows.

To measure the ``hardness'' of candidate items, we use the inner product between the user embedding $\mathbf{u}$ and  embeddings of candidate items in the item pool as the  measurement.
We construct an ANN index based on the current item embedding table to achieve efficient candidate item selection.
Motivated by the hard negative sampling techniques~\cite{kalantidis2020hard}, we use the top $K$ items with the highest similarities to user embeddings as the globally hardest negative samples.
These negative samples usually have very high relevance to user interest, though they are not interacted by the user because they may not be exposed. 
Thus, these samples can be false negatives in recommendation model training and may have huge impacts on model convergence.
To subvert the model learning more effectively, we further propose to replace the original positive samples with pseudo ones obtained by hard sampling. 
We regard the $K$ items with the most dissimilar embeddings to the user embedding as the hardest pseudo positive items.
It is usually difficult for the model to predict high ranking scores for these items, and enforcing the model to match them with dissimilar user interest can greatly impair its ability in modeling the relevance between items and user interest.
A schematic comparison between positive and negative samples on Byzantine clients and benign clients is shown in Fig.~\ref{fig.sam}.
On many benign clients, the positive items are usually similar to user interest, though they may not be the most similar ones.
The negative items are usually randomly selected from local data, which are expected to have low similarities with user interest.
By contrast, on Byzantine clients the negative items are very similar to the user interest, while positive samples are quite diverse from user interest.
Thus, the training samples on Byzantine clients can damage the modeling of user-item relevance, and the poison collaborative signals will propagate in the model training to further affect benign clients.
In addition, the patterns of positive items on Byzantine clients have some similarities to benign users.
For example, if a user has only a few historical interacted items with diverse characteristics, the interacted item may have low similarity to the user interest inferred from previously interacted items~\cite{palomares2018multi}.
Thus, it may be difficult for the server to distinguish Byzantine clients from outlier and cold users (see further discussions in experiments).

After obtaining the hardest positive and negative items, we use the item model to encode their embeddings.
We randomly pair the $K$ negative and pseudo positive items to form $K$ positive-negative sample pairs.
We denote the embedding of a pair of positive and negative items as $\mathbf{i}_p$ and $\mathbf{i}_n$.
A predictor model takes the user embedding and these item embeddings as the input, and calculates the personalized ranking scores for them (independently computed for positive and negative samples).
It can also be implemented by various architectures, such as inner product~\cite{sun2019bert} and neural collaborative filtering~\cite{he2017neural} modules.
The predictor outputs a negative sample score $\hat{y}_n$ and a positive sample score $\hat{y}_p$, which are further used to compute the BPR loss $\mathcal{L}$ as follows:
\begin{equation}
    \mathcal{L}=-\log(\sigma(\hat{y}_p-\hat{y}_n)),
\end{equation}
where $\sigma$ is the sigmoid function.
By optimizing the loss function locally, the malicious clients further compute the malicious gradients by backward propagation.
Both the malicious gradients on Byzantine clients and normal gradients on benign clients are uploaded to the server for aggregation.
The server further distributes the aggregated gradients to each client to conduct local update.
This process is repeated until the model converges.

Our approach addresses the three challenges mentioned in the Introduction Section in the following ways.
First, since globally hardest samples usually have huge impacts on model convergence, the attack can still be effective when there are very limited Byzantine clients.
Second, although the recommendation model may be robust to random noise, it is still vulnerable to our approach because the influence of hard samples is usually accumulated rather than counteracted as noise.
Third, different from many gradient poisoning methods that perturb the entire set of items, our approach only modifies the input samples and their labels.
The local embedding gradients of involved hard positive and negative items are manipulated, while the rest item embeddings are not updated.
The local model updates on Byzantine clients, especially the item embeddings, are still sparse, making it difficult for the server to distinguish between Byzantine and benign clients based on the sparseness of gradients.
Therefore, \textit{FedAttack} can attack federated recommendation effectively and covertly.

%% file: data/experiment.tex
\section{Experiments}\label{sec:Experiments}

\subsection{Datasets and Experimental Settings}

We use two widely used benchmark experiments for recommendation in our experiments.
The first dataset is MovieLens-1M~\cite{harper2015movielens}, which is a movie recommendation dataset.
The second dataset is the Amazon Beauty dataset~\cite{mcauley2015image} (denoted as Beauty).
It is a domain subset of the amazon dump dataset that can be used for e-commerce recommendation.
The statistics of the two datasets are shown in Table~\ref{table.dataset}.
Following~\cite{sun2019bert}, we use the last interacted item of each user as the test sample, the item before the last one for validation, and the rest items for construct training samples.

\begin{table}[h]
\centering
\caption{Statistics of the two datasets.}\label{table.dataset}
\resizebox{0.98\linewidth}{!}{
\begin{tabular}{lrrrrr}
\Xhline{1.0pt}
\multicolumn{1}{c}{Datasets} & \multicolumn{1}{c}{\#Users} & \multicolumn{1}{c}{\#Items} & \multicolumn{1}{c}{\#Actions} & \multicolumn{1}{c}{Avg. length} & \multicolumn{1}{c}{Density} \\ \hline 

MovieLens-1M                        & 6,040                       & 3,416                       & 1.0m                          & 163.5                           & 4.79\%                      \\ 
Beauty                       & 40,226                      & 54,542                      & 0.35m                         & 8.8                             & 0.02\%                      \\
\Xhline{1.0pt}
\end{tabular}
}

%\vspace{-0.1in}
\end{table}

In our experiments, we choose neural collaborative filtering (NCF)~\cite{he2017neural} and BERT4Rec~\cite{sun2019bert} models as the basic victim models for attack.
We use the standard federated learning framework with Adam optimizer~\cite{reddi2020adaptive} for model training, and the learning rate is 1e-3.
The hidden dimension of models is 64.
The number of clients to form a batch is 16.
The maximum epoch is 50.
We randomly sample $M\%$ ($M\in \{1,2,5\}$) of clients as Byzantine clients.
The negative samples on benign clients are randomly drawn  from the whole item set.
The Hit Ratio (HR) and normalized Discounted Cumulative Gain (nDCG)  over the top 5 ranked items are used as the recommendation performance metrics.
Note that the metric is only calculated on benign clients.
Since some recent works figured out that sampled evaluation is biased~\cite{krichene2020sampled}, we evaluate the recommendation performance by ranking the entire item set rather than the sampled subsets based on popularity~\cite{sun2019bert}.
Each experiment is repeated 5 times and the average results are reported.

\begin{table*}[t]
 \caption{Attack performance of different methods under different ratios of Byzantine clients. Lower scores represent better attack effectiveness.} \label{table.performance} 
 \resizebox{0.96\linewidth}{!}{ 
\begin{tabular}{clcccccc|cccccc}
\Xhline{1pt}
\multirow{3}{*}{\textbf{\begin{tabular}[c]{@{}c@{}}Victim\\ Model\end{tabular}}} & \multicolumn{1}{c}{\multirow{3}{*}{\textbf{\begin{tabular}[c]{@{}c@{}}Attack\\ Method\end{tabular}}}} & \multicolumn{6}{c|}{\textbf{MovieLens-1M}}  & \multicolumn{6}{c}{\textbf{Beauty}} \\ \cline{3-14} 
                     & \multicolumn{1}{c}{}                      & \multicolumn{2}{c}{1\%} & \multicolumn{2}{c}{2\%} & \multicolumn{2}{c|}{5\%} & \multicolumn{2}{c}{1\%} & \multicolumn{2}{c}{2\%} & \multicolumn{2}{c}{5\%} \\ \cline{3-14} 
                     & \multicolumn{1}{c}{}                      & HR@5       & nDCG@5     & HR@5       & nDCG@5     & HR@5        & nDCG@5     & HR@5       & nDCG@5     & HR@5       & nDCG@5     & HR@5       & nDCG@5     \\ \hline
\multirow{7}{*}{NCF} & No Attack & 0.0135     & 0.0091     & 0.0135     & 0.0091     & 0.0135      & 0.0091     & 0.0125     & 0.0072     & 0.0125     & 0.0072     & 0.0125     & 0.0072     \\
                     & LabelFlip                                 & 0.0137     & 0.0092     & 0.0133     & 0.0091     & 0.0130      & 0.0090     & 0.0127     & 0.0073     & 0.0122     & 0.0070     & 0.0119     & 0.0069     \\
                     & Gaussian                                  & 0.0138     & 0.0093     & 0.0132     & 0.0091     & 0.0129      & 0.0089     & 0.0125     & 0.0072     & 0.0121     & 0.0070     & 0.0118     & 0.0068     \\
                     & LIE                                       & 0.0133     & 0.0090     & 0.0127     & 0.0089     & 0.0123      & 0.0087     & 0.0120     & 0.0069     & 0.0118     & 0.0068     & 0.0114     & 0.0065     \\
                     & STAT-OPT                                  & 0.0132     & 0.0090     & 0.0125     & 0.0088     & 0.0120      & 0.0086     & 0.0122     & 0.0070     & 0.0119     & 0.0069     & 0.0113     & 0.0065     \\
                     & DYN-OPT                                   & 0.0131     & 0.0090     & 0.0126     & 0.0088     & 0.0122      & 0.0087     & 0.0118     & 0.0068     & 0.0115     & 0.0067     & 0.0110     & 0.0063     \\
                     & FedAttack                                 & 0.0124     & 0.0087     & 0.0116     & 0.0083     & 0.0109      & 0.0079     & 0.0112     & 0.0065     & 0.0107     & 0.0064     & 0.0098     & 0.0058     \\ \hline
\multirow{7}{*}{BERT4Rec}                & No Attack & 0.0550     & 0.0291     & 0.0550     & 0.0291     & 0.0550      & 0.0291     & 0.0159     & 0.0161     & 0.0159     & 0.0161     & 0.0159     & 0.0161     \\
                     & LabelFlip                                 & 0.0544     & 0.0288     & 0.0538     & 0.0285     & 0.0533      & 0.0283     & 0.0155     & 0.0159     & 0.0153     & 0.0157     & 0.0149     & 0.0155     \\
                     & Gaussian                                  & 0.0548     & 0.0289     & 0.0536     & 0.0284     & 0.0532      & 0.0283     & 0.0152     & 0.0156     & 0.0149     & 0.0155     & 0.0144     & 0.0153     \\
                     & LIE                                       & 0.0534     & 0.0283     & 0.0525     & 0.0279     & 0.0513      & 0.0274     & 0.0149     & 0.0154     & 0.0144     & 0.0152     & 0.0140     & 0.0150     \\
                     & STAT-OPT                                  & 0.0536     & 0.0285     & 0.0521     & 0.0277     & 0.0510      & 0.0272     & 0.0150     & 0.0155     & 0.0146     & 0.0153     & 0.0141     & 0.0151     \\
                     & DYN-OPT                                   & 0.0530     & 0.0280     & 0.0517     & 0.0272     & 0.0501      & 0.0268     & 0.0146     & 0.0152     & 0.0144     & 0.0151     & 0.0139     & 0.0150     \\
                     & FedAttack                                 & 0.0511     & 0.0270     & 0.0484     & 0.0258     & 0.0459      & 0.0245     & 0.0137     & 0.0149     & 0.0132     & 0.0146     & 0.0122     & 0.0139     \\ \Xhline{1pt}
\end{tabular}
}
\end{table*}

\begin{figure*}[!t]
	\centering
	\includegraphics[width=0.96\linewidth]{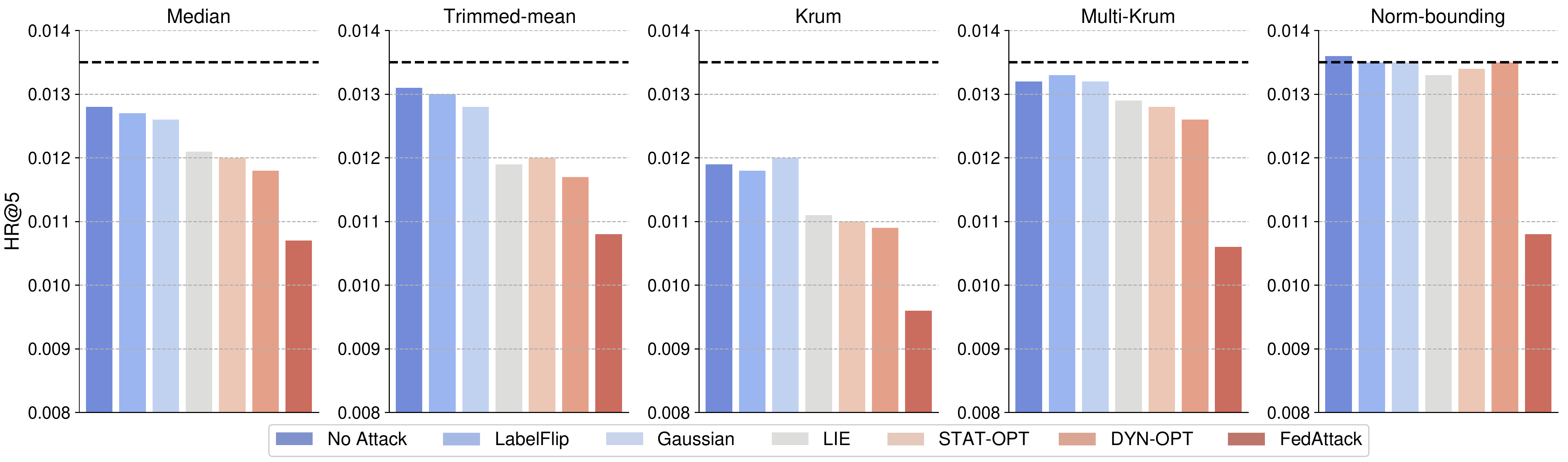}
	
\caption{Performance of different attack methods under different defense mechanisms. The black dashed lines represent the performance without attack.}\label{fig.ex1}
\end{figure*}

\subsection{Attack Performance Evaluation}

We compare the attack performance of \textit{FedAttack} with several untargeted poisoning methods on federated learning, including:
(1) \textit{LabelFlip}~\cite{fang2020local}, flip the positive and negative item labels on Byzantine clients;
(2) \textit{Gaussian}~\cite{fang2020local}, using Gaussian noise that has the same distribution with the before-attack local models to replace Byzantine clients' local models;
(3) \textit{LIE}~\cite{baruch2019little}, adding small amounts of noise to  the average of benign gradients;
(4) \textit{STAT-OPT}~\cite{fang2020local}, adding constant noise with opposite directions to the average of benign gradients;
(5) \textit{DYN-OPT}~\cite{shejwalkar2021manipulating}, adding noise with dynamic malicious directions to the average of benign gradients.
Since the total number of clients in federated recommender systems is usually large, we limit the ratio of malicious clients to three different small values, i.e., 1\%, 2\%, and 5\%.
In addition, we report the model results without attack for benchmarking the attack performance. 
The results are shown in Table~\ref{table.performance}, from which we have several findings.
First, all compared attack methods can degrade the recommendation performance when there are relatively sufficient Byzantine clients (e.g., 5\%), while most baseline methods are ineffective when Byzantine clients are very limited.
Especially, the \textit{LabelFlip} and \textit{Gaussian} methods can even slightly raise the performance when there are only 1\% malicious clients. 
This may be because these perturbation methods can make the model more robust to the noise in user behaviors.
Second, gradient poisoning methods such as \textit{LIE}, \textit{STAT-OPT}, and \textit{DYN-OPT} have better attack effectiveness than \textit{LabelFlip} and \textit{Gaussian}.
This is because the noise introduced by perturbing the labels or the models on a small fraction of clients is  weak, which cannot effectively subvert the training process.
Third, \textit{FedAttack} has consistently better attack performance than other baselines, and their gaps are significant ($p<0.01$ in t-test).
This is because the hardest positive and negative samples have strong impacts on model convergence, and the poisoned collaborative information can be further propagated to benign clients in model training.
In addition, the advantage of our approach in terms of the relative performance degradation can usually be larger when a smaller percentage of Byzantine clients is used.
The results show that hard sampling can be a powerful  technique for untargeted federated recommendation poisoning with limited malicious clients.

\begin{figure}[!t]
	\centering
\subfigure[Impacts of malicious gradient detection on attack performance. The black dashed line is the recommendation performance without attack.]{
	\includegraphics[width=0.8\linewidth]{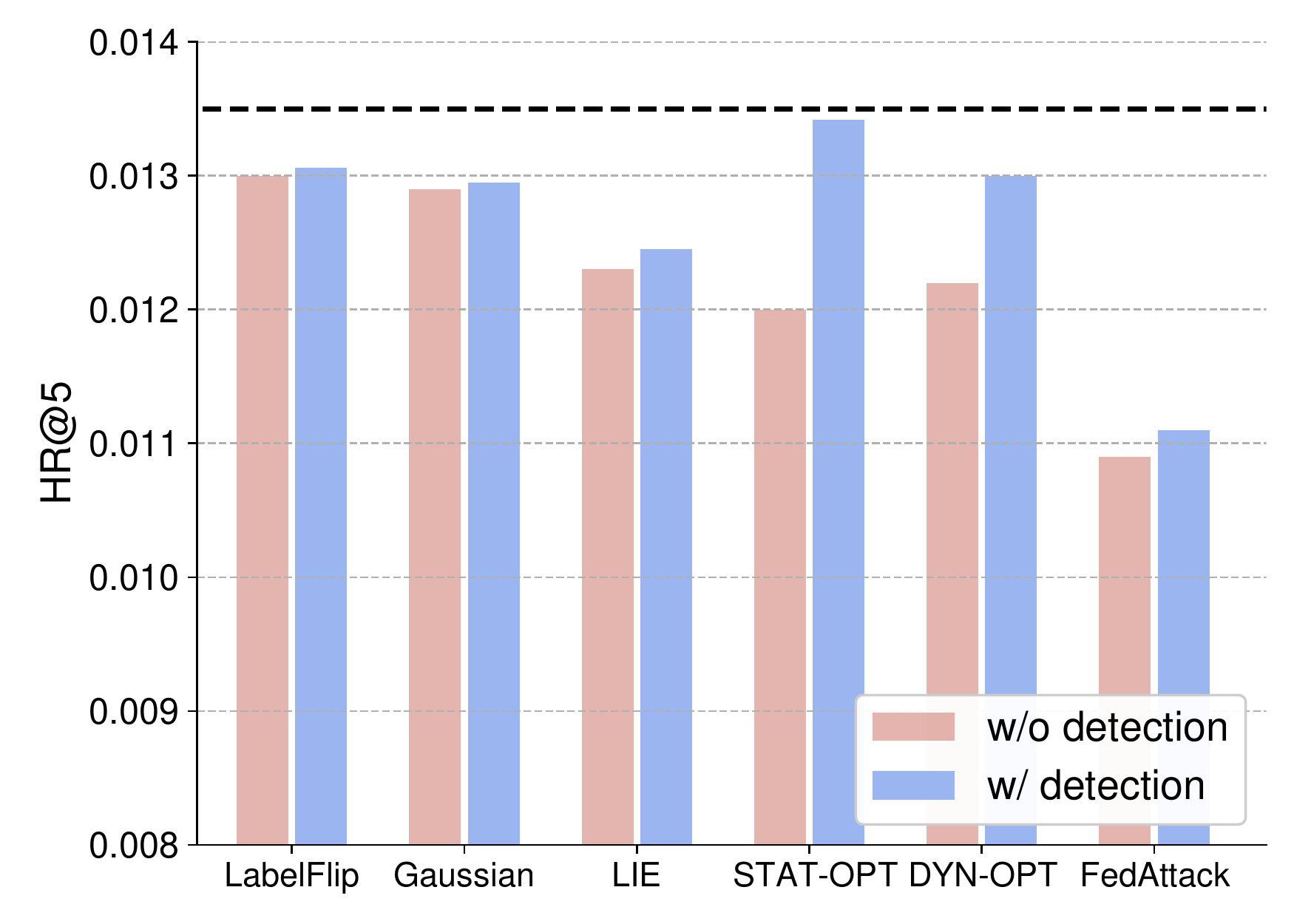}
	}
	\subfigure[Malicious gradient detection accuracy.]{
	\includegraphics[width=0.8\linewidth]{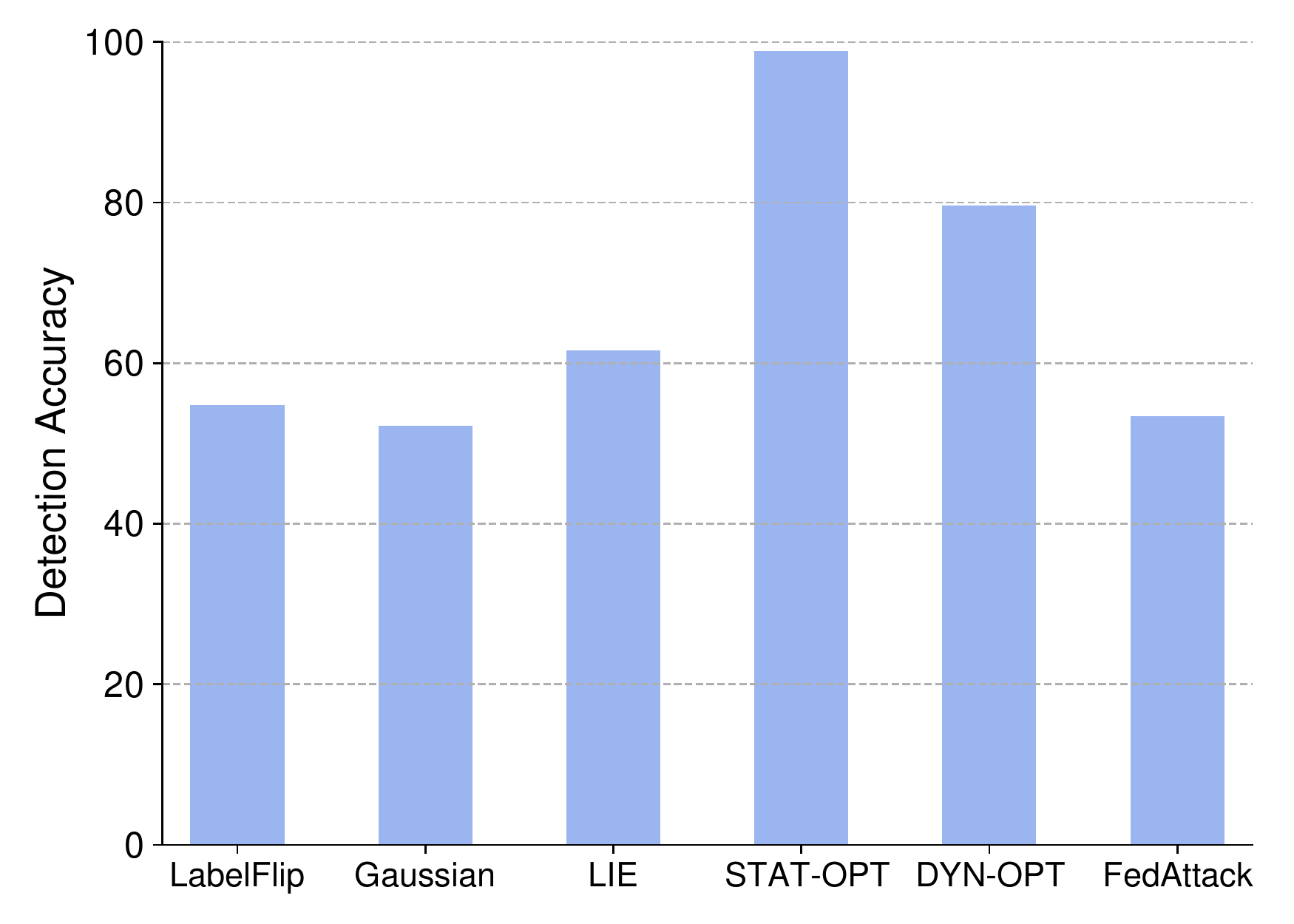}
	}
\caption{Results of malicious gradient detection.}\label{fig.ex2}
\end{figure}

\begin{figure}[!t]
	\centering
	\includegraphics[width=0.8\linewidth]{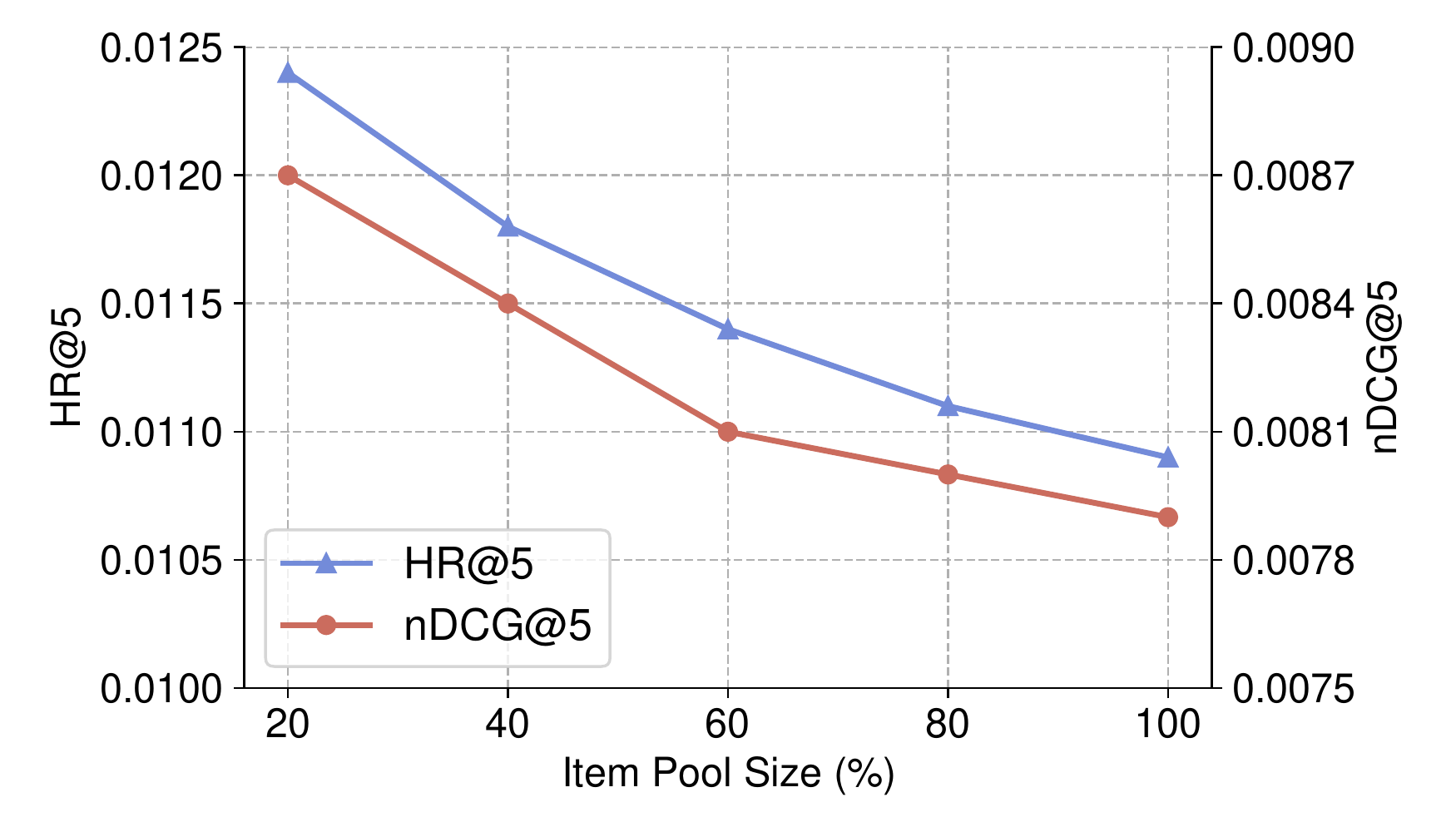}
	
\caption{Influence of candidate item pool size.}\label{fig.ex3}
\end{figure}

\subsection{Attack Effectiveness Under Defense}

To verify the effectiveness of \textit{FedAttack} in circumventing the defense of robust aggregation mechanisms in FL, we compare the attack performance of different methods when the server uses the following  strategies:
(1) \textit{Median}~\cite{yin2018byzantine},  using the median value of received gradients for each dimension;
(2) \textit{Trimmed-mean}~\cite{yin2018byzantine}, removing some largest and smallest values for each dimension and using the average of the rest;
(3) \textit{Krum}~\cite{blanchard2017machine}, selecting the local model that is the most similar to other models as the global model;
(4) \textit{Multi-Krum}~\cite{blanchard2017machine}, iteratively selecting several local models through Krum and aggregating them;
(5) \textit{Norm-bounding}~\cite{sun2019can}, using a threshold (we use 2 in our experiments) to clip the norm of gradients.
We use 5\% Byzantine clients in this and the later experiments.
The results on the MovieLens-1M dataset are shown in Fig.~\ref{fig.ex1} (the results on Beauty show similar trends and are omitted due to space limit).
We find that some defense mechanisms such as \textit{Median}, \textit{Trimmed-mean} and \textit{Krum} have huge impacts on recommendation accuracy, and their performance losses can even be larger than sufferring from some attack methods without defense.
Thus, they may not be good choices for defending untargeted poisoning attacks on federated recommendation.
\textit{Mutli-Krum} and \textit{Norm-bound} do not substantially hurt the recommendation accuracy, but only \textit{Norm-bounding} can fully defend poisoning attack baselines.
This is because \textit{Mutli-Krum} may incorrectly filter some normal gradients, while norm-bounding usually does not hurt the training accuracy if the threshold is appropriate.
\textit{FedAttack} cannot be effectively defended by all these robust aggregation methods.
This is because it only modifies training samples, and the Byzantine users also have some similar patterns with benign users.
Thus, \textit{FedAttack} can circumvent their defense and effectively degrade the model performance.

\begin{figure}[!t]
	\centering
	\includegraphics[width=0.99\linewidth]{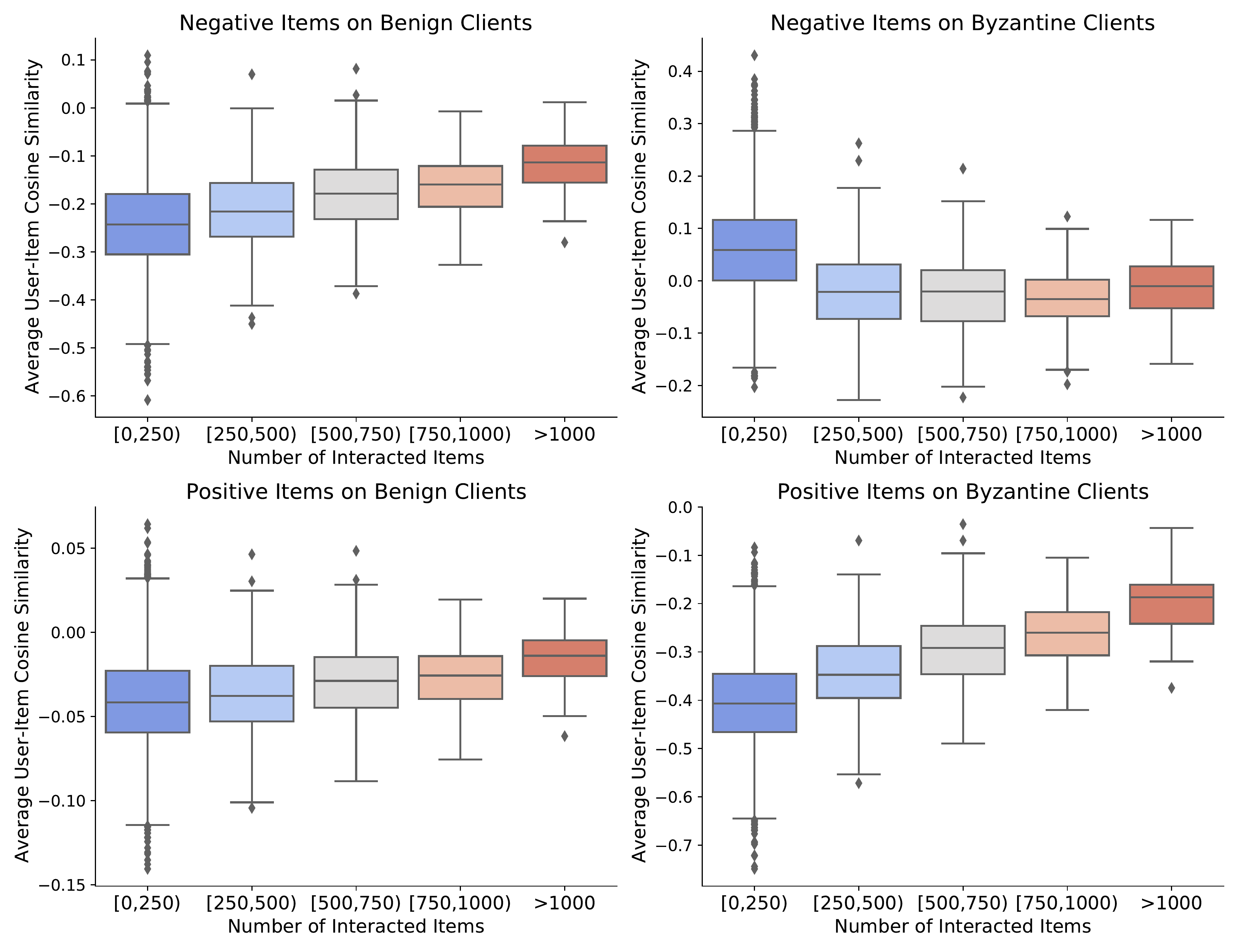}
	
\caption{The average similarity between positive/negative items and users w.r.t. different numbers of interacted items on benign and Byzantine clients.}\label{fig.ex4}
\end{figure}

\begin{figure*}[!t]
	\centering
	\includegraphics[width=0.98\linewidth]{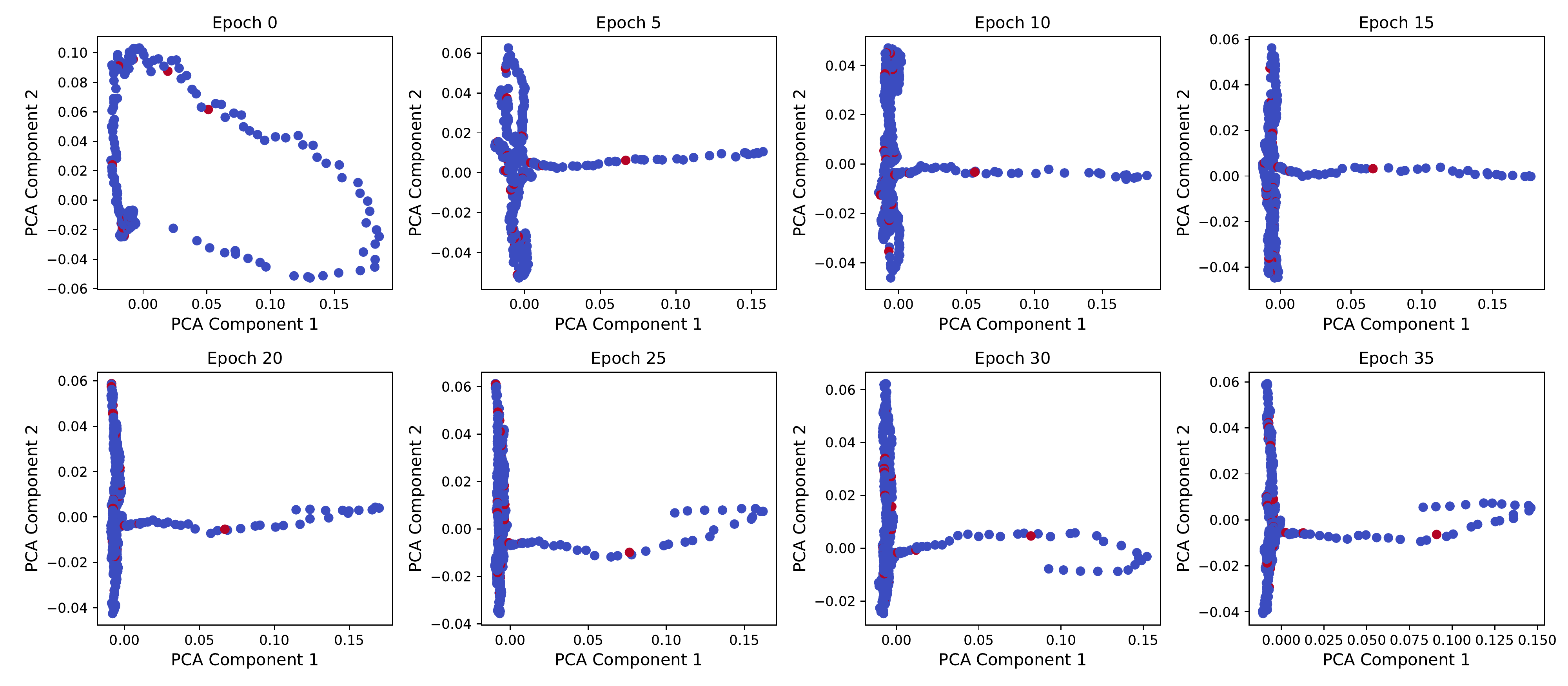}
	
\caption{Visualization of local gradients in different epochs of model training. Red and blue points represent gradients from Byzantine and benign clients, respectively.}\label{fig.ex5}
\end{figure*}

\subsection{Influence of Malicious Gradient Detection}

To verify whether \textit{FedAttack} is still effective when there are strong malicious detection mechanisms on the server, we study an extreme case where the server  detects and filters potential malicious gradients based on strong background knowledge of the attacker.\footnote{We do not consider client detection because the client identity may not be exposed to the server in federated learning~\cite{liu2021flame}.}
We assume that the server has some samples of malicious gradients for training an anomaly detection model.
To simulate this scenario, we first run an experiment to collect malicious gradients.
We concatenate the average gradients and the gradients of a client to be classified as the input of a multi-layer feedforward network for detection.
Note that we down-sample the normal gradient samples to form a balanced binary classification.
After obtaining the detection model, we re-run the federated recommendation learning and use it to detect potential malicious gradients.
The server excludes the positively detected gradients before aggregation.
The attack performance of different methods on MovieLens-1M and the server's detection accuracy on them are shown in Fig.~\ref{fig.ex2} (the results on Beauty show similar patterns and are omitted).
We find that the attack performance of all methods decreases to different extents when there are detection mechanisms, which is intuitive because some malicious clients are filtered.
We can see gradient poisoning methods (especially STAT-OPT) are easier to be detected.
This is because their gradients usually have similar patterns, and their norm and variance are also larger than normal gradients.
In addition, we have an interesting finding that \textit{LabelFlip} and \textit{Gaussian} attacks are difficult to detect, which is different from the observations in general FL~\cite{taheri2020defending}.
This may be because the user interaction data is quite noisy, and it is hard to distinguish between the patterns of these perturbations and noise.
Besides, our \textit{FedAttack} approach can achieve the best attack performance under strong supervised detection mechanisms.
It can hardly be detected by the server even the server has prior knowledge of the malicious gradients.
Thus, \textit{FedAttack} is even more difficult to be  perceived by the server in real-world federated recommendation model training.

\subsection{Influence of Item Pool Size}

In practice the attacker may not know the entire item set, and the candidate item pool is usually a subset. 
To study the influence of item pool size on the attack performance, we compare the results of \textit{FedAttack}  under different sizes of the item pool. 
We use randomly sampled subsets of the whole item set to simulate the scenarios when the attacker has partial knowledge of the full item set.
The results on MovieLens-1M are shown in Fig.~\ref{fig.ex3} (similar patterns are observed on Beauty).
We find that the poisoning effectiveness of \textit{FedAttack} is still satisfactory when a relatively large item subset is used (e.g., 60\%), but it is suboptimal when the item pool is too small.
This is because the hardest samples in a small subset may not be sufficiently effective to subvert the model training.
Thus, the attacker needs to crawl a relatively large item pool to ensure that the hard samples are representative enough.

\subsection{Sample Hardness Analysis}

To explore whether the hardness of samples on Byzantine clients has significant differences with benign clients, we compare the average similarity between the embeddings of positive or negative items and the embeddings of users with different numbers of interacted items, as shown in Fig.~\ref{fig.ex4}.
For users with fewer interacted items, both positive and negative samples on benign clients have low average similarities to user embedding, while the similarity variance is large.
This is because the interest inferred from the profiles of less active users can be too inaccurate or not comprehensive to predict future behaviors.
Different from benign clients, most negative samples on Byzantine clients have higher similarities to user interest while most positive samples are irrelevant to user interest.
In addition, we can see that the hardness distribution of positive samples on Byzantine clients with many interacted items has many overlaps with benign but inactive clients. 
It indicates the behaviors of Byzantine clients have similar patterns with some benign cold users whose behaviors are diverse and noisy.
Since the server does not know the raw behavior data distribution on local clients due to privacy reasons, it cannot effectively distinguish between Byzantine and benign clients merely based on their gradients (also see the next section).
Thus, \textit{FedAttack} can subvert the model training in a concealed way.

\subsection{Gradient Visualization}

Finally, we use PCA to visualize the gradients on benign clients and Byzantine clients to see whether they are visually distinguishable, as shown in Fig.~\ref{fig.ex5}.
From all subplots, we find that malicious gradients have similar PCA component distributions with normal gradients.
Thus, it is very difficult to distinguish between normal and malicious gradients.
In addition, we find a very interesting phenomenon that there are many outlier gradients during model training and most of them are normal ones.
This phenomenon reflects many outlier gradients from benign clients may be mistakenly filtered by defense or detection mechanism, which may heavily impair the recommendation performance on these users.
These visualization results further verify the covertness of \textit{FedAttack} and the difficulty in defending it poisoning.

%% file: data/conclusion.tex
\section{Conclusion}\label{sec:Conclusion}

In this paper, we present an effective untargeted attack method for poisoning federated recommender systems, named \textit{FedAttack}.
We propose to employ hard negative sampling as to poison federated recommendation by selecting positive and negative candidate items on Byzantine clients that can subvert the model training.
More specifically, we select items that are most similar to the user interest inferred from local user profiles as negative samples, while regarding items that are most dissimilar to user interest as positive samples.
The Byzantine clients train the recommendation model on these poisoned samples, and generate malicious gradients to upload to the server for aggregation.
Extensive experiments on two benchmark datasets for recommendation validate \textit{FedAttack} can degrade recommendation performance more effectively than many baseline methods.
In addition, it can effectively circumvent the existing defense and detection mechanisms for federated learning.
Our work shows existing federated recommendation methods can be vulnerable, and the design of future federated recommender systems should be more secure and robust to these attacks, which is one of our future work.